\documentclass[final]{svjour2}
\usepackage{graphicx}
\usepackage{rotating}
\usepackage{amssymb}
\usepackage{mathptmx}
\usepackage[square, numbers]{natbib}
\usepackage{caption}
\setcitestyle{square}
\makeatletter
\journalname{Journal of Low Temperature Physics}

\bibpunct{[}{]}{,}{n}{}{,}

\begin{document}

\newcommand{\hdblarrow}{H\makebox[0.9ex][l]{$\downdownarrows$}-}
\title{Characterizing Atacama B-mode Search Detectors with a Half-Wave Plate}

\author{S. M. Simon\textsuperscript{\dag} \and J. W. Appel\textsuperscript{\ddag} \and L. E. Campusano\textsuperscript{$\cap$} \and S. K. Choi\textsuperscript{\dag}\and K. T. Crowley\textsuperscript{\dag}  \and T. Essinger-Hileman\textsuperscript{\ddag} \and P. Gallardo\textsuperscript{$\star$} \and S. P. Ho\textsuperscript{\dag} \and A. Kusaka\textsuperscript{$\diamond$}\textsuperscript{\dag} \and F. Nati \textsuperscript{$\circ$} \and G.A. Palma\textsuperscript{$\ast$} \and L. A. Page\textsuperscript{\dag} \and S. Raghunathan\textsuperscript{$\cap$}\and S. T. Staggs \textsuperscript{\dag}}

\institute{\textsuperscript{\dag}Department of Physics, Princeton University\\  Princeton, NJ 08544, USA\\
\email{smstwo@princeton.edu}\\ \textsuperscript{\ddag}Department of Physics and Astronomy, Johns Hopkins University\\  Baltimore, MD 21218, USA\\ \textsuperscript{$\cap$}Departamento de Astronomia, Universidad de Chile\\ \textsuperscript{$\ast$}Departamento de Fisica, Universidad de Chile\\  FCFM, Santiago, Chile\\ \textsuperscript{$\star$}Department of Physics, Cornell University\\  Ithaca, NY 14853, USA\\ \textsuperscript{$\diamond$}Physics Division, Lawrence Berkeley National Lab\\ Berkeley, CA 94720, USA\\ \textsuperscript{$\circ$}Department of Physics and Astronomy, University of Pennsylvania\\Philadelphia, PA 19104}


\maketitle

\begin{abstract}
The Atacama B-Mode Search (ABS) instrument is a cryogenic ($\sim$10~K) crossed-Dragone telescope located at an elevation of 5190 m in the Atacama Desert in Chile that observed for three seasons between February 2012 and October 2014. ABS observed the Cosmic Microwave Background (CMB) at large angular scales ($40<\ell<500$) to limit the B-mode polarization spectrum around the primordial B-mode peak from inflationary gravity waves at $\ell \sim100$. The ABS focal plane consists of 480 transition-edge sensor (TES) bolometers. They are coupled to orthogonal polarizations from a planar ortho-mode transducer (OMT) and observe at 145 GHz. ABS employs an ambient-temperature, rapidly rotating half-wave plate (HWP) to mitigate systematic effects and move the signal band away from atmospheric $1/f$ noise, allowing for the recovery of large angular scales. We discuss how the signal at the second harmonic of the HWP rotation frequency can be used for data selection and for monitoring the detector responsivities.

\keywords{Atacama B-mode Search, Cosmic Microwave Background, polarization, half-wave plate}

\end{abstract}

\section{Introduction}
Thomson scattering at recombination linearly polarizes the CMB. This polarization can be decomposed into even (E-mode) and odd (B-mode) parity components. An inflationary period in the early universe might have produced gravitational waves, which produce both E-modes and B-modes at recombination. Scalar density perturbations can also source E-modes but not B-modes, so the detection of a primordial B-mode signal would be clear evidence for inflation. ABS observed the CMB on large angular scales for three seasons from February 2012 to October 2014 with the aim of measuring or limiting the B-mode spectrum from multipole moments of $\ell \approx 40$ to $\ell \approx 500$, a range that includes the primordial B-mode peak at $\ell\sim100$ whose amplitude would give the energy scale of inflation \cite{kami1,seljak}

\section{The Atacama B-mode Search Instrument}
The ABS instrument is a crossed-Dragone telescope located at an elevation of 5190 m in the Parque Astron\'{o}mico in Chile. The primary ABS field is a $\sim$2400~square degree, low-foreground patch below the Galactic plane centered at (RA, DEC) = (25$^\circ$,-42$^\circ$). ABS has an angular resolution of 32$'$ full width at half maximum and an array NEQ\footnote{Noise equivalent sensitivity to a single linear Stokes parameter (Q or U)} of $\sim$30~$\mu$K$\sqrt{\mathrm{s}}$. The telescope employs 60~cm mirrors and a 25~cm aperture stop cooled to 4~K by a system of two pulse-tube coolers to reduce loading and increase sensitivity \cite{Tom}. The focal plane is cooled to a base temperature of 300 mK by a \textsuperscript{3}He/\textsuperscript{4}He adsorption refrigerator system \cite{Penn}.

The ABS focal plane array contains 240 pixels (480 detectors) that are designed for operation at 145~GHz and fabricated at the National Institute of Standards and Technology \cite{Tom_ltd,Yoon,John}. The detectors are feedhorn-coupled TES bolometers with on-chip band-defining filters. Each pixel has two MoCu TES bolometers coupled to orthogonal polarizations from the OMT \cite{Yoon}. The TESes are kept stable via negative electrothermal feedback and are read out with time-domain multiplexing. The two halves of the ABS array were fabricated in two batches, batch A and batch B. Due to unexpected shifts in the dielectric constant between fabrications, batch B has a bandpass shifted up by $\sim15$~GHz.

The ABS HWP is an ambient-temperature 330~mm diameter $\alpha$-cut sapphire plate that is 3.15~mm thick with 305~$\mu$m anti-reflection (AR) coatings of Rogers RT/Duroid 6002 fluoropolymer laminate \cite{Tom}. The HWP is rotated continuously on air bearings at $f=2.55$~Hz, which modulates the incoming polarization in the bolometer timestreams at $4f$. The modulation is good for controlling systematics, eliminates the need for differencing timestreams from orthogonal pairs of detectors to gain polarization sensitivity, and reduces the sensitivity loss associated with filtering the timestreams. Additionally, the ABS HWP is the first optical element of the telescope, which allows the separation of instrumental polarization and signal polarization. The HWP modulation permits the measurement of celestial polarization at frequencies well above the $1/f$ knee ($\sim$1~Hz) of the atmosphere. The median knee frequency of the detectors after demodulation is 2.0~mHz, which allows for the recovery of information from large angular scales \cite{Kusaka}. The HWP also enables the characterization of the ABS instrument in new ways, which we describe in the following.

\section{Second Harmonic of the HWP}
The detector timestream is composed of the unpolarized sky intensity, the modulated polarization signal, white noise, and spurious modulation signals $A(\chi)$ that depend on the HWP angle $\chi$; $A(\chi)$ consists of components at every harmonic $n$ of the HWP rotation frequency \cite{Kusaka}. The $n=2$ component, $A_{2}(\chi)$, is a sinusoidal signal at the second harmonic of the HWP rotation frequency. This $2f$ signal has a contribution from differential emissivity along the two crystal axes of the sapphire HWP and an additional contribution from differential transmission through the HWP. One source of differential transmission arises because the loss tangents along the two crystal axes of the sapphire are different. The second contribution is a result of differential reflection due to the different indices of refraction for the two axes. The signal due to differential transmission scales linearly with sky temperature and thus with precipitable water vapor (PWV) for low PWV. The loading from the PWV changes as a function of elevation $\theta$ as $1/\sin{\theta}$. In the ABS observations of the primary field, the elevation varies by less than 3$^{\circ}$, which corresponds to a change in loading of less than $5\%$. Thus, we take loading from PWV as constant in this analysis. We can then write $A_{2}(\chi)$ as:
\begin{equation} \label{eq:2f}
A_{2}(\chi)=A_{2c}\cos{(2\chi)}+A_{2s}\sin{(2\chi)},
\end{equation}
where $A_{2c}=(c_0 +c_1 PWV)$ and $A_{2s}=(s_0 +s_1 PWV)$, and $c_0$, $c_1$, $s_0$, and $s_1$ are constants. For our analysis, we use the measured PWV values from the Atacama Pathfinder EXperiment (APEX) radiometer. The $2f$ signal is clean, simple, and, to a good approximation, independent of the demodulation process for CMB analysis since the polarization modulation occurs at $4f$. This makes the $2f$ signal an ideal calibration source.

\subsection{Data Selection}
\begin{figure}[h]
\centering
\includegraphics[width=0.69\textwidth]{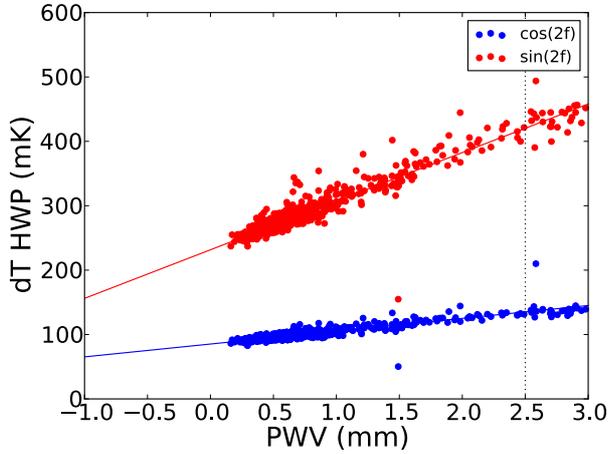}
\caption{The $A_{2c}$ (blue) and $A_{2s}$ (red) signals for a single bolometer are plotted above as a function of PWV with their fits. Each CES contributes a point to both the $A_{2c}$ and $A_{2s}$ distributions. The vertical dotted line indicates the 2.5 mm PWV fitting cutoff. The slopes of the $A_{2c}$ and $A_{2s}$ signals depend on the polarization angle of the detector. (Color figure online.)}
\label{reg}
\end{figure}

\begin{figure}[h]
\centering
\includegraphics[width=0.85\textwidth]{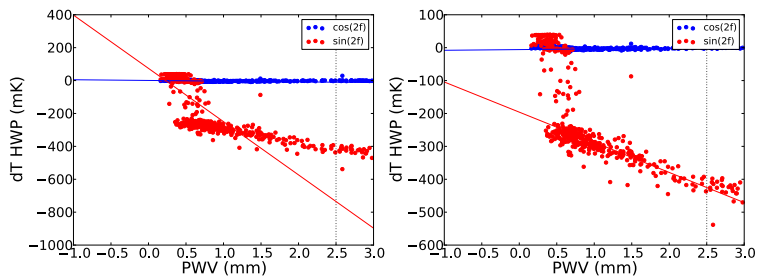}
\caption{One of the $\lesssim10$ type 1 detectors is shown before ($left$) and after ($right$) reprocessing with the regions of poor biasing excluded from the fit. (Color figure online.)}
\label{flag1}
\end{figure}

The $2f$ signal can be used to determine if the detectors are biased and responding properly. For each bolometer, the $A_{2c}$ and $A_{2s}$ signals are characterized with respect to PWV for each $\sim$hour-long constant elevation scan (CES) for each sub-season of data as in Fig.~\ref{reg}. Due to discrete changes in responsivity over time that are discussed in the next section, the data for the first two seasons of observation are broken up into four sub-seasons for which the full analysis is repeated. The $A_{2c}$ and $A_{2s}$ distributions are linearly fit for data with PWV values below 2.5 mm to exclude possible nonlinearities in detector behavior at high loading. To remove outliers, we perform a preliminary fit of the data and then use only the inner 85$\%$ of the residuals to determine the optimal fit. The standard deviation of the residuals is determined for the $A_{2c}$ and $A_{2s}$ distributions for each detector, and detectors with zero response are cut. In each sub-season, there are five to seven detectors that do not operate properly at low PWV due to being improperly biased. These detectors are flagged as ``type 1" detectors and reprocessed as in Fig.~\ref{flag1}. The standard deviations of the $A_{2c}$ and $A_{2s}$ distributions ($\sigma_{cos}$ and $\sigma_{sin}$ respectively) are used to make a statistical parameter $\chi_{CES}^2$:
\begin{equation} \label{eq:chi}
\chi_{CES}^2=\left(\frac{A_{2c, CES}-A_{2c, fit}}{\sigma_{cos}}\right)^2 +\left(\frac{A_{2s, CES}-A_{2s, fit}}{\sigma_{sin}}\right)^2  ,
\end{equation}
where $A_{2c, CES}$ and $A_{2s, CES}$ are the amplitudes of the cosine and sine components for a given CES and $A_{2c, fit}$ and $A_{2s, fit}$ are the amplitudes from the fit at the given PWV of the CES. Using the distributions of $\chi_{CES}^2$ for each detector and the entire detector distribution, we determine the optimal $\chi_{CES}^2$ cutting threshold for each sub-season. For all seasons, if $\chi_{CES}^2>36.0$, the detector timestream for the given CES is cut. The reprocessed type 1 detectors in the second season require tighter $\chi_{CES}^2$ constraints ($\chi_{CES}^2>20.25$) than nominal detectors, so all reprocessed detectors remain flagged so that we can apply a separate variable cutting parameter.

Additionally, the linear $A_{2c}$ and $A_{2s}$ fits from this data selection technique can be used to recover the PWV for CESes where there are no APEX radiometer measurements ($\sim8\%$ of the total data). To determine the PWV, we take the median value of the calculated PWV from a subset of 100 detectors that are highly sensitive to the variation in PWV and are consistently well behaved throughout all seasons. This method allows for a small number of the 100 detectors used to be ill-behaved without affecting the PWV estimation. Once calculated, the PWV can then be used to reject ill-behaved detectors that were not used in the PWV calculation.

\subsection{Responsivity}
\begin{figure}[h]
\centering
\includegraphics[width=0.75\textwidth]{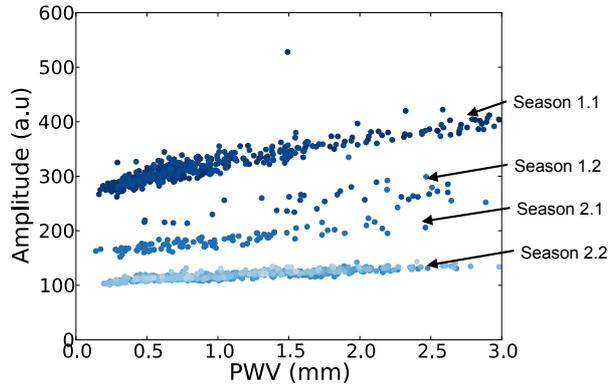}
\caption{The $2f$ amplitude as a function of PWV for seasons 1 and 2 is plotted above for a single batch B detector. The points range from early (dark blue) to late (light blue) times. The responsivity decreases discretely in four distinct epochs. (Color figure online.)}
\label{resp}
\end{figure}

In addition to data selection, the amplitude of the $2f$ signal can also be used to track the instrument responsivity over the three seasons of observations. The $2f$ signal as defined in Eq.~(\ref{eq:2f}) can be expressed in terms of an amplitude linear with PWV and a constant phase. Using the $2f$ amplitude data, we define four time epochs of constant responsivity within the first two seasons of data. The duration of each epoch is typically a few months, although the shortest one is two weeks. While the responsivities of batch A detectors remain constant between epochs, the $2f$ amplitudes of batch B detectors decay discretely between the four epochs of constant responsivity (Fig.~\ref{resp}). Seasons 1 and 2 each have two stable epochs. The discrete decays in responsivity of batch B detectors are consistent between both the slope and $y$-intercept of the $2f$ amplitude. The $y$-intercept of the $2f$ amplitude for each epoch is used as a measure of the responsivity of a detector in that epoch relative to a given reference epoch. While we only use the $2f$ amplitude to track the responsivity between epochs in our responsivity model, it can also be used to determine the relative responsivity of the detectors with respect to a reference detector. When this is done, the responsivity decay from the $2f$ amplitude between epochs for batch B detectors is consistent with that found in relative responsivity measurements performed with a polarizing wire grid \cite{Tajima}. We have not yet identified the source of the variability, but note that each epoch of stable batch B responsivity was preceded by a month or longer in which no observations were made and the cryostat warmed to ambient temperature, after which the cryostat was re-pumped to reach vacuum and cooled before resuming observations. 

\section{Summary}
ABS observed the CMB for three seasons from the Atacama Desert and pioneered the use of an ambient-temperature, continuously-rotating HWP for ground-based CMB observations. In addition to allowing for the mitigation of systematic effects and reducing the instrument's $1/f$ noise, the ABS HWP has provided new methods for instrument and detector characterization and calibration. We have demonstrated that the $2f$ signal is ideal for calibration and can be used for data selection, to recover missing PWV values, and to track detector responsivity changes.

\begin{acknowledgements}
This work is supported by a NASA Space Technology Research Fellowship. Work at NIST is supported by the NIST Innovations in Measurement Science program. Work at Princeton University is supported by the NSF through awards PHY-0355328 and PHY-085587, NASA through award NNX08AE03G, and Wilkinson Misrahi funds. PWV measurements were provided by APEX. SR acknowledges his CONICYT PhD studentship, CONICYT Anillo project (ACT No. 1122), and the Aspen Center for Physics, which is supported by National Science Foundation grant PHY-1066293. A. K. was supported by a Dicke Fellowship.  This work was supported in part by the U.S. Department of Energy under contract No. DE-AC02-05CH11231.\end{acknowledgements}

\end{document}